\documentclass[letterpaper]{jpconf}
\usepackage{epsfig}
\begin{document}
\title{Production of the $h_c$ and $h_b$ and Implications for 
Quarkonium Spectroscopy}

\author{Stephen Godfrey}

\address{Ottawa-Carleton Institute for Physics, \\
Department of Physics, Carleton University, Ottawa, Canada K1S 5B6}

\ead{godfrey@physics.carleton.ca}

\begin{abstract}
The recent observation of the $h_c$ is an important test of 
QCD calculations and provides constraints on models of quarkonium 
spectroscopy.  In this contribution I discuss some of these 
implications and describe methods to search for the $h_c$ and $h_b$ 
via radiative transitions and other means.

\end{abstract}.

\section{Introduction}

Over the years there have been numerous calculations of quarkonia
spectra.  On the one hand, first principles calculations starting with 
the QCD Lagrangian such as Lattice QCD and NRQCD provide a rigorous 
test of the theory while on the other hand, quark models can provide 
more intuitive insights into these systems and provide important 
phenomenological guidance towards their study \cite{Brambilla:2004wf}.  
In both cases it is 
absolutely necessary to test theory against experiment. 
The $P$-wave singlet 
quarkonium states are particularly significant as they are the first 
place we can really test our understanding of the spin-spin 
interaction between quarks where  
complications due to relativistic and other effects 
are less important than in light quark mesons.
In this short writeup I will summarize 
some of the different predictions for the $^1P_1-^3P_{cog}$ splittings
\footnote{Where cog stands for the triplet J=0, 1, 2 centre of 
gravity.}. 
We will see that the recent CLEO \cite{Tomaradze:2004sk}
and E835 \cite{Patrignani:2004nf} measurements of 
the $h_c$ mass provide an important test of theoretical 
predictions.  I will also briefly describe alternative ways of 
searching for the $h_c$ and $h_b$ \cite{Barnes:2004uc}
via radiative transitions 
\cite{Godfrey:2002rp}
and in $B$-meson decays (for the $h_c$)
\cite{Eichten:2002qv,Suzuki:2002sq,Gu:2002nh,Colangelo:2003sa}.
 
\section{$\Delta(M(^1P_1) - M(^3P_{cog}))$ as a Test of Quarkonium 
Calculations}

There are numerous calculations of quarkonia  
properties \cite{Brambilla:2004wf}.  
The measurement of the singlet-triplet splitting 
is an important validation of lattice QCD calculations and pNRQCD 
calculations.  It is also an important means of testing various models.
For example, in the quark model the triplet-singlet splitting tests 
the Lorentz nature of the confining potential.  The standard 
Lorentz vector 1-gluon-exchange at short distance with a Lorentz 
scalar confining potential gives a very short range spin-spin 
interaction.  In contrast, a Lorentz vector confining potential 
implies a long-range interaction.  
Representative predictions for the $M(^1P_1) - M(^3P_{cog})$ splitting 
are summarized in Fig.~1. A more complete listing is
given in Ref.~\cite{Godfrey:2002rp}. 

In quark potential models the 1-gluon-exchange spin-spin 
interaction is described by:
\begin{equation}
H_{q\bar{q}}^{hyp}={{32\pi} \over 9} {{\alpha_s}\over{m_q 
m_{\bar{q}}}} \vec{S}_q \cdot \vec{S}_{\bar{q}} \; \delta^3 (\vec{r})
\end{equation}
The $\delta$-function is short range but will be smeared out by 
relativistic effects.  The Godfrey-Isgur quark model \cite{Godfrey:1985xj}
smeares the $\delta$-function with
a Gaussian and predicts $M(^3P_{cog})>M(^1P_1)$.  In contrast, 
McClary and Byers \cite{McClary:1983xw}
include spin-independent relativistic corrections 
and find $M(^3P_{cog})<M(^1P_1)$. Finally, Franzini \cite{Franzini:1992nk}
includes a 
Lorentz vector confining potential and finds $M(^3P_{cog})<M(^1P_1)$ 
with a large splitting. 

Pantaleone and Tye \cite{Pantaleone:1987qh}
calculated the splitting using perturbative QCD and 
also found a small splitting with $M(^3P_{cog})<M(^1P_1)$ but noted 
that other contributions  such as relativisitic corrections and 
coupled channel effects could alter this result.  Lattice QCD 
finds $M(^3P_{cog})>M(^1P_1)$ but with large errors \cite{Manke:2000dg}.  
Ultimately 
LQCD will provide the definetive result but more precise results are 
needed.

The point of these examples is that there is a wide variation in the 
predictions.  There is a strong need for experimental data to test these 
results.

\begin{figure}[t]
\epsfig{file=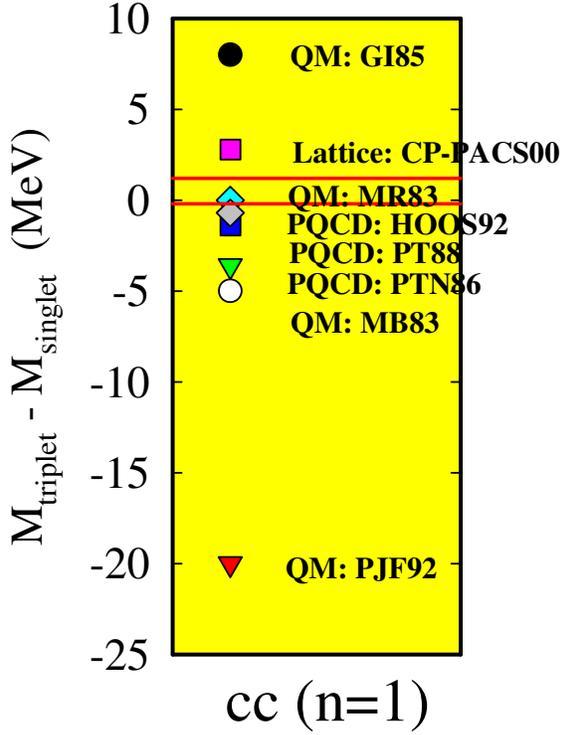,width=18pc,clip=}
\hspace{2pc}%
\begin{minipage}[b]{18pc}
\caption{Comparison of the measured and predicted 
$1^3P_{cog}-1^1P_1 (c\bar{c})$ mass splitting. The horizontal lines 
show the 1-sigma bounds using the CLEO $h_c$ mass
measurement \cite{Tomaradze:2004sk}.
The theoretical predictions correspond to: 
GI85 \cite{Godfrey:1985xj}, 
CP-PACS00 \cite{Manke:2000dg},
MR83 \cite{Moxhay:1983vu},
HOOS92 \cite{Halzen:1992wm},
PT88 \cite{Pantaleone:1987qh},
PTN86 \cite{Pantaleone:1985uf},
MB83 \cite{McClary:1983xw},
PJF92 \cite{Franzini:1992nk}.
 \\ \qquad \\ \qquad}
\end{minipage}
\end{figure}

\section{Production of Singlet $P$-wave States}

There are a number of ways to produce and detect the singlet 
$P$-wave states.  The $h_c$ was recently observed in the reaction 
$\psi'\to \pi^0 h_c \to (\gamma\gamma)(\gamma \eta_c)$ by the CLEO 
collaboration \cite{Tomaradze:2004sk} 
and a less convincing signal was seen
in $\bar{p}p \to h_c \to \eta_c \gamma$ by E835 at 
FNAL \cite{Patrignani:2004nf}.  
It has been suggested that the singlet $P$-waves states 
could also be produced in the radiative cascades 
$n^3S_1\stackrel{M1}{\to} {n'} {^1S_0} +\gamma 
\stackrel{E1}{\to} (1^1P_1) +\gamma\gamma $ \cite{Godfrey:2002rp}
and in $B$-meson decay, $B\to h_c + X$ 
\cite{Eichten:2002qv,Suzuki:2002sq,Gu:2002nh,Colangelo:2003sa}.

In all cases the radiative decay $h_{c,b} \to 
\eta_{c,b} + \gamma$ results in a clean final state.  To estimate the 
BR requires knowing all important partial decay widths.
The E1 width for the $h_c$ is given by \cite{Godfrey:2002rp}
\begin{equation}
\Gamma[h_c(^1P_1) \to \eta_c(^1S_0) + \gamma] = \frac{4}{9} \alpha \; e_Q^2 \; 
\omega^3 \; |\langle ^1S_0 | r | ^1P_1 \rangle |^2 =354 \hbox{ keV}
\end{equation}
where $\alpha = 1/137.036$ is the fine-structure constant, $e_Q$ is the
quark charge in units of $|e|$ ($2/3$ for $Q=c$ and $-1/3$ for $Q=b$), 
and $\omega$ is the photon's energy. The overlap integrals were 
obtained using the wavefunctions of Ref. \cite{Godfrey:1985xj}.
The strong widths are estimated to be 
\cite{Godfrey:2002rp,Kwong:1987ak,Maltoni:2000km}
\begin{equation}
\Gamma[h_c(^1P_1)\to {\rm hadrons}]={5\over{2n_f}} \times 
\Gamma[\chi_{c1}(^3P_1) \to {\rm hadrons}]=533  \hbox{ keV}
\end{equation}
\begin{equation}
\Gamma[h_c(^1P_1) \to gg+\gamma ] = {{36}\over 5} e_q^2 {\alpha \over 
\alpha_s } \Gamma[h_c(^1P_1) \to ggg] = 52 \hbox{ keV}
\end{equation}
where $n_f$ is the number of light quark flavours in the final state 
which we will take to be 3.  
For $b\bar{b}$ we combined the theoretical estimates 
for the radiative transitions 
$\chi_{b1}(^3P_1) \to \Upsilon({^3S_1}) \gamma$ with the measured 
BR's \cite{Eidelman:2004wy} 
to estimate the $\chi_{b1}$ hadronic width \cite{Godfrey:2002rp}.
For the $h_c$, 
${\cal B}[h_c \to \eta_c +\gamma]=37.7\%$ 
and for the $h_b$, ${\cal B}[h_b \to \eta_b +\gamma]=41.4\%$.

It was pointed out long ago that a promising way to produce the 
$h_c$ (and $h_b$) is via the decay $\psi'(2S) \to h_c +\pi^0$
(and $\Upsilon(3S) \to h_b \pi^0 $) 
\cite{Voloshin:1985em,Kuang:1988bz,Ko:1994nw,Kuang:2002hz}.  
Estimates 
for the branching ratio are ${\cal B}[\psi' \to h_c + \pi^0]=0.1-0.3\% $
\cite{Voloshin:1985em,Kuang:1988bz,Ko:1994nw,Kuang:2002hz}.  
Combining this result 
with the predicted BR for 
${\cal B}(h_c \to \eta_c +\gamma)$ gives 
${\cal B}(\psi'\to h_c \pi^0)\times
{\cal B}( h_c\to \eta_c \gamma)\simeq 3.8\times 10^{-4}$ 
which would yield roughly 400 events for $10^6$ $\psi'$s 
produced.  Likewise, we obtain 
${\cal B}(\Upsilon(3S) \to h_b \pi^0)\times  
{\cal B}( h_b\to \eta_b \gamma )\simeq 4\times 
10^{-4}$ which also yields  $\sim 400$ events for $10^6$ 
$\Upsilon(3S)$s.  
Kuang  and Yan \cite{Kuang:1981se}
have also considered the related spin-flip transition 
$\Upsilon(3S)\to h_b(^1P_1)+\pi\pi$ 
which may provide an additional path to the $h_b$.

CLEO recently observed the $h_c$ in $\psi' \to h_c +\pi^0$ \cite{Tomaradze:2004sk}.  
They measured
${\cal B}(\psi' \to \pi^0 h_c) \times {\cal B}(h_c\to \gamma \eta_c) 
=(2-6)\times 10^{-4}$. This is in good agreement with the theoretical 
prediction and is an important validation of the 
$\psi' \to h_c + \pi^0$ calculations.

Another possibility for producing the singlet $P$-wave 
$c\bar{c}$ and $b\bar{b}$ states is via 
electromagnetic cascades \cite{Godfrey:2002rp} such as;
\begin{equation}
\psi(2S)\stackrel{M1}{\to} \eta_c(2S) +\gamma 
\stackrel{E1}{\to} h_c(1P) +\gamma\gamma \stackrel{E1}{\to} \eta_c (1S)
+\gamma\gamma\gamma .
\end{equation}
As before, we need the BR's to estimate the expected number of events. 
The M1 transition widths are given by:
\begin{equation}
\Gamma[\psi' (2^3S_1) \to \eta_c'(2^1S_0) + \gamma] = 
{{4 \alpha e_Q^2}\over{ 3 m_Q^2}}
\omega^3 |\langle f | j_0 (kr/2) | i \rangle |^2 =0.051 \hbox{ keV}
\end{equation}
where we take $m_c=1.628$~GeV and as before use the wavefunctions of
Ref. \cite{Godfrey:1985xj}.
Using the measured $\psi'$ width gives 
${\cal B}[\psi' (2^3S_1) \to \eta_c'(2^1S_0) + \gamma] = 0.018\% $.  For the 
next decay in the chain, an E1 transition, we estimate \cite{Godfrey:2002rp}:
\begin{equation} 
\Gamma[\eta_c'(2^1S_0) \to h_c(1^1P_1) + \gamma] 
= \frac{4}{3} \alpha \; e_Q^2 \; 
\omega^3 \; |\langle ^1P_1 | r | ^1S_0 \rangle |^2 = 51.3 \hbox{ keV}.
\end{equation}
and
\begin{equation}
\Gamma(^1S_0\to gg)={{27\pi}\over{5(\pi^2-9) }} {1\over{\alpha_s}}
\times \Gamma(^3S_1\to ggg)=7.4 \hbox{ MeV}
\end{equation}
where we haven't shown the QCD corrections that were included in 
obtaining this result.  This results in $BR(\eta_c'\to h_c \gamma)=0.69 \% $. 
Combining this result with the $\psi'\to \eta_c' \gamma$ BR we obtain
${\cal B}(\psi' \to \eta_c' \gamma)\times  
{\cal B}(\eta_c'\to h_c \gamma) \simeq 10^{-6}$ 
which would yield only 1 event per $10^6$ $\psi'$'s.  Similarly, we find 
${\cal B}(\Upsilon(3S)\to \eta_b' \gamma)\times 
{\cal B} (\eta_b' \to h_b \gamma)=2.6\times 10^{-7}$ resulting in only 0.3 
events per $10^6 \; \Upsilon(3S)$'s.  In a gross understatement, this 
would be quite the challenge for experimentalists.

The final possibility is to produce the $h_c$ in $B$ decay.  This mode has 
been explored in a number of papers 
\cite{Eichten:2002qv,Suzuki:2002sq,Gu:2002nh,Colangelo:2003sa}
and is supported by the Belle observation \cite{belle02} 
of the $\eta_c(2S)$ in $B\to \eta_c(2S) K$.  Combining the estimate of 
${\cal B}[B\to h_c + K]\simeq 0.1\%$ 
\cite{Eichten:2002qv,Suzuki:2002sq,Gu:2002nh}
with the BR for $h_c\to \eta_c \gamma$ 
we obtain 
${\cal B}(B\to h_c +K)\times {\cal B}(h_c \to \eta_c \gamma) 
\sim 4 \times 10^{-4}$.
Belle and Babar should be able to observe this.

\section{Summary}

In the last decade there has been considerable theoretical progress, 
especially in lattice QCD.  We need comparable experimental results to 
check these calculations.  Recently the $h_c$ has been observed 
which provides an important reality check for theory. The $h_c$ is 
found to be almost degenerate with the $1^3P_{cog}(c\bar{c})$ 
and referring 
to Fig.~1 we see that the new measurements rule out a Lorentz vector 
confining potential,  and demonstrate the need for improved models and 
for improved calculations using PQCD.

\section*{Acknowledgments}
This work was supported by the Natural Sciences and 
Engineering Research Council of Canada

\end{document}